\documentclass[12pt]{article}               
\usepackage{a4}
\usepackage{epsfig}
\usepackage{amsmath,amssymb}
\begin{document}
\begin{flushright} 
\vspace*{-30mm}
{\bf LAL 04-20}\\
May 2004\\
\end{flushright}
\vspace{15mm}\begin{center}
{\Large\bf Experimental Implications for a Linear Collider}\\
{\Large\bf   of the SUSY Dark Matter Scenario}\\
\vspace{10mm}

{\sc P.~Bambade$^a$, M.~Berggren$^b$, F.~Richard$^a$, 
  Z.~Zhang$^{a,}$\footnote{E-mail: bambade@lal.in2p3.fr, 
  mikael.berggren@cern.ch, richard@lal.in2p3.fr, zhangzq@lal.in2p3.fr}}\\
\vspace{5mm}
{\bf $^a$Laboratoire de l'Acc\'el\'erateur Lin\'eaire,}\\
 IN2P3-CNRS et Universit\'e de Paris-Sud, B\^at.\ 200, BP 34, 
 91898 Orsay Cedex, France\\
\vspace{3mm}
{\bf $^b$Laboratoire de Physique Nucl\'eaire
et de Hautes Energies,} \\ 
IN2P3-CNRS et Universit\'e Paris VI-VII, 4 place Jussieu, 
Tour 33 - Rez de chauss\'ee,\\
75252 Paris Cedex 05, France\\
\vspace{10mm}
\end{center}
\vskip 4 cm
\noindent
\centerline {\it Work presented at the International Conference on 
Linear Colliders (LCWS04)}\\
\centerline {\it 19-23 April 2004, ``Le Carr\'e des Sciences'', Paris, France}
\par
\vskip 4 cm
\vspace{5mm}
\begin{abstract}
 This paper presents the detection issues for the lightest 
 slepton $\tilde{\tau}_1$ at a future $e^+e^-$ TeV collider given the dark 
 matter constraints set on the SUSY mass spectrum by the WMAP results. It 
 intends to illustrate the importance of an optimal detection of energetic 
 electrons in the very forward region for an efficient rejection of the 
 $\gamma\gamma$ background. The TESLA parameters have been used in the case 
 of head-on collisions and in the case of a $10$\,mrad half crossing angle. 
\end{abstract}

\vspace{10mm}
\vfil\eject
\baselineskip=13.07pt

\section{Introduction}
The present paper is motivated by the increasing awareness in the community of 
the role of an $e^+e^-$ Linear Collider (LC) for a precise determination of 
the SUSY parameters which are needed to interpret the dark matter (DM) content
of the universe. After the WMAP results leading to an accuracy on 
$\Omega_{\rm DM}h^2$ at the $10\%$ level and awaiting for the Planck 
mission in $2007$ which aims at $2\%$, it seems appropriate to check that 
a LC can do its job properly on this essential topic. 
The main issue is to compare the SUSY prediction derived from 
the collider measurement and the DM result observed in our universe.
A significant mismatch would reveal the existence of extra components of DM 
in the universe which are predicted within SUSY (e.g.\ the gravitinos) or 
beyond SUSY (e.g.\ the axions). 

An important issue is also to check with this type of physics the effect of 
some choices discussed for the future LC.  
This paper will explain in which way the DM issue can provide some useful 
informations. After an introduction of these arguments intended for 
`pedestrians', a quantitative study will be presented to illustrate 
the problem. 

In the SUSY scenario with R-Parity conservation, the lightest SUSY particle 
(LSP) is the lightest neutralino $\chi$. This particle is considered as 
the best candidate to satisfy the cosmological constraints on DM in 
the universe. DM constraints have been recently re-examined~\cite{ellismap} 
within the mSUGRA scenario, confronting the precise predictions obtained 
after the WMAP results. These data imply, for many of the working points 
retained, a very small difference between the lightest slepton mass, 
the SUSY partner of the $\tau$ which will be called $\tilde{\tau}_1$,
and the LSP mass since one of the preferred mechanism to regulate
the amount of DM in the universe is the so-called `co-annihilation mechanism'. 
Since this feature is quite general and goes beyond the mSUGRA scheme as 
pointed out in \cite{mad}, one should investigate the possible experimental 
consequences on the detection of sleptons at a LC. The detectability of 
the tau slepton in such a small mass difference has been discussed recently 
in \cite{arnow} in mSUGRA for a $500$\,GeV LC collider.

To understand the effect of mass degeneracy, one should recall the mechanism 
which regulates the amount of DM in the universe and is based 
on thermodynamics and on annihilation cross sections 
between SUSY particles. The $\chi\chi$ annihilation cross section is small 
since it occurs in a p-wave (Fermi exclusion principle for identical fermions)
with a kinematic suppression at threshold. A way out is to annihilate 
$\chi$ with $\tilde{\tau}_1$ but, given the Boltzmann law, this can only occur
during the cooling of the early universe if these particles have a small mass 
difference. How small should this mass difference be? Typically the ratio 
between the stau and neutralino populations is given by $\exp(-\Delta m/T_f)$ 
where $\Delta m$ is the mass difference and where $T_f$ is the freeze-out 
temperature which is $\sim m/20$. To avoid a strong depopulation of 
$\tilde{\tau}_1$, the mass difference should therefore be below $m/20$ which 
in practice means typically below $10$\,GeV.
Since $\tilde{\tau}_1$ is significantly lighter than the other sleptons, 
the amount of DM will therefore primarily depend on the mass difference 
between $\chi$ and $\tilde{\tau}_1$ masses. 
When the $\chi$ mass increases one needs to increase the annihilation 
efficiency to keep at the same level the amount of DM density in the universe.
This means that one needs to reduce even further $\Delta m$. 
When the neutralino mass reaches about $500$\,GeV this mechanism does not work 
anymore which gives an important upper limit on SUSY particles. 

Admittedly there are other ways to solve this problem as will be discussed 
in detail in the next section. 

\vspace{3mm}
From this qualitative presentation, one therefore concludes that the SUSY 
scenario implies that: 
\begin{enumerate}
\item the mass difference between $\tilde{\tau}_1$ and the LSP is 
      likely to be below $10$\,GeV,
\item the amount of DM depends critically on the mass of the stau 
particle. 
\end{enumerate}
Experimentally this means that the stau channel should be cleanly detected 
to measure its mass through a threshold scan. Near threshold, 
the cross section is at the $10$\,fb level with a potentially very large 
background due to the four fermion final states, the so-called 
`$\gamma\gamma$' background, which is at the nb level. 
Moreover $\tilde{\tau}_1$ decays into a $\tau$ lepton with one or two 
neutrinos in the final state which even further reduces the amount of visible 
energy. 

In usual cases the standard backgrounds can be eliminated by requiring that 
the two observed leptons be acoplanar thus eliminating the $\gamma\gamma$ 
background provided that the forward veto forces spectator electrons to be 
emitted at almost zero angle in the four fermion process. 
This veto usually starts above a polar angle of $5$\,mrad which is sufficient 
to cope with ordinary SUSY mass differences for a $500$\,GeV collider. 
The present paper intends to quantify the effect for the SUSY solutions 
retained in \cite{ellismap}. 

The next issue to be considered is the reduction in efficiency of this veto 
in case there is a crossing angle between electron and positron beams,
$\pm 10$\,mrad, as needed in the warm technology and as also envisaged in 
the TESLA scheme. 

Finally, one should also worry about the efficiency of this veto 
given the very large overlaid background produced by beam-beam interaction 
which hits the very forward electromagnetic calorimeter. This paper therefore 
also intends to provide an input on these various aspects.     

\section{DM and SUSY} \label{sec:dm_susy}
\begin{figure}[htb]
\vskip 5mm
\begin{center}
\psfig{figure=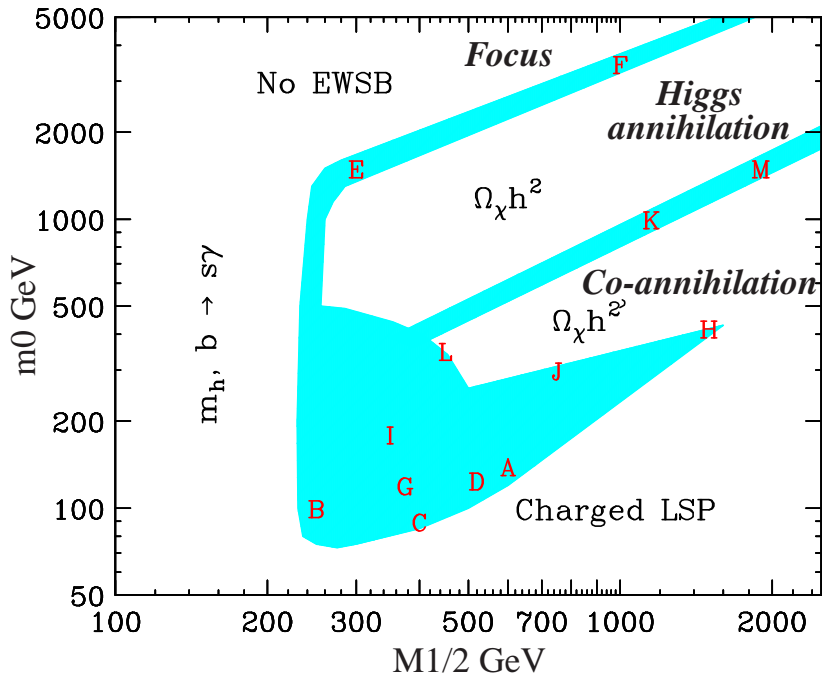,width=11truecm }
\end{center}
\vskip -5mm
\caption{\sl \label{fig:para_space}
 Schematic view of the various DM solutions described in the text 
 and display of the various working points proposed in \cite{ellismap}. }
\end{figure}

For a simple discussion of the issues presented in the introduction, 
the mSUGRA scheme is hereafter adopted. The mass spectrum depends on two 
parameters $m0$ and $M1/2$, the common masses of scalars and gauginos 
superpartners at the unification scale. The parameter $\mu$, 
defining the higgsino mass, is derived, in absolute value, by imposing 
the electroweak symmetry breaking (EWSB) condition in terms of these 
two parameters and of tan$\beta$, the ratio of the vacuum expectations 
which appear in the two Higgs doublets of SUSY. The various regions allowed 
by EWSB and DM constraints and the position of the various working points 
chosen are schematically displayed in figure~\ref{fig:para_space}.
\begin{enumerate}
\item For moderate values of $m0$ and $M1/2$, the LSP solution is a Bino, 
 the SUSY partner of the SM $U(1)$ gauge boson. 
 The annihilation cross section for a pair of LSP, which controls the amount 
 of DM in the universe, proceeds through selectron exchange and 
 can be easily adjusted to cope with the WMAP constraints.
\item For larger values of $M1/2$ and moderate values of $m0$, the rate of 
 annihilation through selectron exchange becomes insufficient and 
 co-annihilation between the LSP and the lightest sleptons, 
 in particular $\tilde{\tau}_1$, are needed. 
 When the LSP mass increases, the co-annihilation process should increase, 
 meaning that the mass difference between the LSP and the slepton should 
 tend to zero at some point. For moderate values of tan$\beta$ 
 (not larger than $30$), this maximum value is about $500$\,GeV, 
 which incidentally means that this type of mechanism for generating DM 
 should be covered by a TeV collider.  The exact amount of DM in the universe 
 depends crucially on the mass of the LSP, which can be easily determined 
 using end-points smuon (or selectron) decays as can be seen in 
 figure~\ref{fig:emu}, and on the mass of $\tilde{\tau}_1$ which 
 can be determined through a threshold scan.\footnote{An alternative 
 approach would be to analyze the high energy spectrum. 
 This method however requires that the stau mass is relatively small 
 with respect to the beam energy and the signal production cross section 
 is large enough in comparison with background contributions in the final 
 energy spectrum.} 
 If the mass difference between the lightest stau and the LSP is too small, 
 the separation from the background will become problematic and these scans 
 could be extremely time consuming or even impossible. 
 The accuracy on the $\tilde{\tau}_1$ mass is primordial since it governs 
 the accuracy on the prediction of DM in the universe which will be known
 to a few percent after the Planck mission. This aspect will therefore 
 be the main emphasis of the present study.
\item For large values of $m0$ and $M1/2$ there are, within mSUGRA, 
 two scenarios to cope with the WMAP constraints.

 In the `focus' solution, $m0$ is large but $\mu$ can be smaller than $M1/2$ 
 meaning that the LSP is of the higgsino type and therefore can annihilate 
 into $WW/ZZ$ pairs in the early universe. This solution can lead to 
 a degeneracy between the LSP and the lightest chargino but, given that 
 the chargino cross section is much larger than the slepton one, 
 this situation can be dealt with using events having photons
 originating from initial state radiation (ISR) as was shown at LEP2.
 It turns out, however, as will be discussed in section~\ref{sec:focus}, 
 that the degree of degeneracy is not severe for light neutralinos and 
 therefore one can achieve standard accuracies for light gaugino masses.
 As pointed out in reference \cite{baer} and as can be seen in 
 figure~\ref{fig:baer10},\footnote{This is a modified version of figure 8
 from reference \cite{baer} by exchanging the $x$ and $y$ axes so that
 they are the same as in figures~\ref{fig:para_space} and \ref{fig:tanx10}.}
 this type of solution is not well covered by LHC since, with very heavy 
 squarks and gluinos, the only accessible SUSY particles are the first 
 generation chargino and the two lightest neutralinos which need to be 
 produced directly through quark-antiquark annihilation.
\begin{figure}[htb]
\begin{center}
\psfig{figure=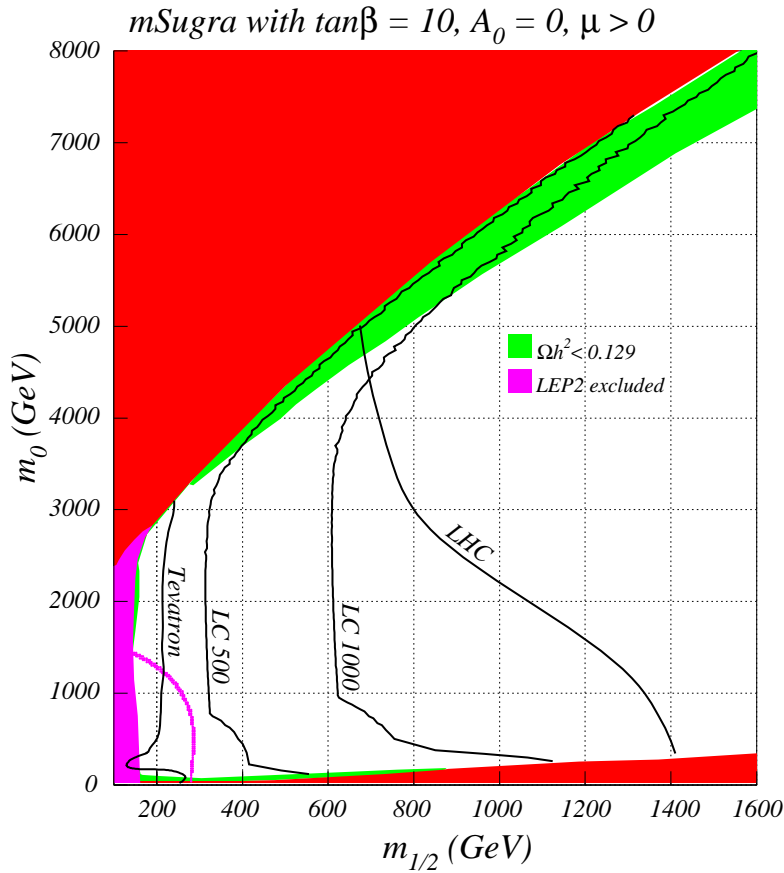,width=11truecm }
\end{center}
\vskip -15mm
\caption{\sl \label{fig:baer10}
 A modified version of figure 8 from reference \cite{baer} showing
 the allowed DM solutions in green in the mSUGRA scheme for tan$\beta=10$ 
 and $\mu>0$. The regions to the left of the black curves 
 are the domain covered by LHC and a LC at $500$ and $1000$\,GeV respectively
 for an integrated luminosity of $100$\,fb$^{-1}$ ($10$\,fb$^{-1}$ for the
 Tevatron).}
\end{figure} 

In the `Higgs annihilation' solution, the LSP mass is close to half the heavy 
CP odd Higgs mass meaning that LSP annihilate through $s-$channel into 
a heavy Higgs. This mechanism only operates for very large values of 
$\tan\beta$, typically above $40$, which are allowed and even needed to 
accomplish unification between the Yukawa couplings of the third generation 
(but discarded in some `string inspired' theories~\cite{ellistr}). 
This solution leads to no particular constraints on the detection of the LSP 
but corresponds to mass solution which tends to fall beyond the reach of LC 
and, in some cases, even also beyond the reach of LHC.
\end{enumerate}
Figure~\ref{fig:baer10} shows, for moderate $\tan\beta$, 
the allowed region in green and the expected coverage from LHC and LC. 
In the co-annihilation region, at moderate $m0$, the LHC has a wider coverage  but the LC can fully cover the DM solutions provided that there are no detection problems.
In the focus region, LC extends the reach of LHC but, 
as previously mentioned, the detection issue is less critical and 
will be briefly discussed in section~\ref{sec:focus}.

The $g-2$ measurement~\cite{update} indicates a deviation with respect to 
the SM prediction based on $e^+e^-$ data. The deviation favors $\mu>0$ and 
moderate mSUGRA masses. The indication is confirmed since the $e^+e^-$ 
CMD2 data used
are found in good agreement with the recent data~\cite{KLOE} of KLOE 
based on the radiative return method. 
The discrepancy~\cite{davier} with the $\tau$ decay analysis 
from LEP1 and CLEO, which is significant, remains however to be clarified.

Figure~\ref{fig:tanx10} from \cite{likely} indicates that, 
taking into account the old $g-2$ constraint, 
the most likely solutions correspond to the co-annihilation region. 
With the recent update from BNL~\cite{update}, the deviation with 
respect to the $e^+e^-$ SM prediction, is almost reaching 3 
standard deviations (s.d.)\ and should therefore increase the significance 
of this indication. 
\begin{figure}[htb]
\begin{center}
\psfig{figure=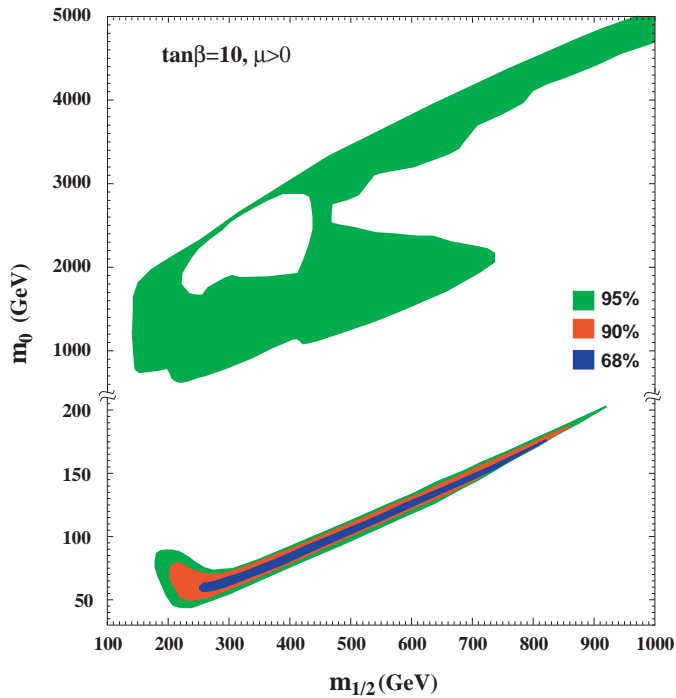,width=9truecm }
\end{center}
\vskip -5mm
\caption{\sl \label{fig:tanx10}
 For $\tan\beta=10$ and $\mu>0$, the acceptable mSUGRA solutions~\cite{likely} 
 are shown respecting the various constraints including the $g-2$ results. 
 The most likely region, with $68\%$ likelihood, corresponds to the 
 co-annihilation region.}
\end{figure}

There are several caveats to above conclusions.
First, the co-annihilation solution may sound heavily `fine tuned' 
since it requires a very tight correlation between the two mSUGRA mass 
parameters. One can however notice that in the so called `gaugino mediated 
SUSY breaking' scenario, $m0$ which is loop-mediated is naturally small and 
a mass degeneracy between the neutralino and the right-handed sleptons occurs
naturally~\cite{gaugino}. 
The two other mechanisms previously discussed are also fine tuned and 
one cannot really object to any of these solutions in the absence of 
a definite SUSY breaking scheme. One may also argue that, 
in a general MSSM scheme, there is more flexibility and one could 
for instance have a wino-like LSP, in which case there is no need 
for the co-annihilation mechanism to regulate primordial DM. 
In this general MSSM approach and for other phenomenological reasons 
(e.g.\ CP violation in the flavor sector, proton lifetime constraints) 
one may wish to have very heavy sleptons at least for the first two 
generations. In this case the LSP mass could not be anymore determined 
from the smuon analysis but would require to use the chargino/neutralino 
channels. There could however still be a similar experimental problem, 
with co-annihilation between the stau and the neutralino with small mass 
difference.  

Table~\ref{tab:scenarios} recalls the mSUGRA solutions retained in 
\cite{ellismap} and 
clearly indicates the trend described above. Figure~\ref{fig:emu} shows
the energy distribution of the muons originating from smuon decays 
for solution D$^\prime$. 
Two end points are clearly visible and their measurement allows to precisely 
extract the values of the mass of the slepton and of the LSP provided 
the lower point can be separated from the  $\gamma\gamma$ background. 
The serious concern is about the stau analysis which, as shown in the same 
figure, has very soft final state particles to deal with.
\begin{table}[htb]
\caption{\sl \label{tab:scenarios}
 Working points (model) taken from reference \cite{ellismap}. Note that in 
 some cases (E$^\prime$,F$^\prime$,H$^\prime$,M$^\prime$) the resulting DM 
 content obtained from Micromegas~\cite{micro} does not match the WMAP 
 constraints.}
\begin{center}
\footnotesize
\begin{tabular}{|c|c|c|c|c|c|c|c|c|c|c|c|c|c|}
\hline 
Model & A$^\prime$ & B$^\prime$ & C$^\prime$ & D$^\prime$ & E$^\prime$ & F$^\prime$ & G$^\prime$ & H$^\prime$ & I$^\prime$ & J$^\prime$ & K$^\prime$ & L$^\prime$ & M$^\prime$  \\
\hline
\hline 
 $M1/2$ & $600$ & $250$ & $400$ & $525$ & $300$ & $1000$ & $375$ & 
 $935$ & $350$ & $750$ & $1300$ & $450$ & $1840$ \\
\hline 
 $m0$ & $107$ & $57$ & $80$ & $101$ & $1532$ & $3440$ & $113$ & $244$ & 
 $181$ & $299$ & $1001$ & $303$ & $1125$ \\   
\hline 
 $\tan\beta$ & $5$ & $10$ & $10$ & $10$ & $10$ & $10$ & $20$ & $20$ & $35$ & 
 $35$ & $46$ & $47$ & $51$ \\  
\hline 
 $\mu(m_Z)$ & $773$ & $339$ & $519$ & $-663$ & $217$ & $606$ & $485$ & 
 $1092$ & $452$ & $891$ & $-1420$ & $563$ & $1940$ \\
\hline 
 $m_\chi$ & $242$ & $95$ & $158$ & $212$ & $112$ & $421$ & $148$ & $388$ & 
 $138$ & $309$ & $554$ & $181$ & $794$ \\
\hline 
 $m_{e_R}$ & $251$ & $117$ & $174$ & $224$ & $1534$ & $3454$ & $185$ & 
 $426$ & $227$ & $410$ & $1109$ & $348$ & $1312$ \\
\hline 
 $m_{\tau_1}$ & $249$ & $109$ & $167$ & $217$ & $1521$ & $3427$ & $157$ & 
 $391$ & $150$ & $312$ & $896$ & $194$ & $796$ \\
\hline
\hline 
 $\Delta m$ & $7$ & $14$ & $9$ & $5$ & $1409$ & $3006$ & $9$ & 
 $3$ & $12$ & $3$ & $342$ & $13$ & $2$ \\
\hline 
\hline 
 $\Omega_{\rm DM}h^2$ & $0.09$ & $0.12$ & $0.12$ & $0.09$ & $0.33$ & 
 $2.56$ & $0.12$ & $0.16$ & $0.12$ & $0.08$ & $0.12$ & $0.11$ & $0.27$ \\  
\hline
\end{tabular}
\end{center}
\end{table}  

\section{Forward set-up used to veto electrons} \label{sec:fwd}
\begin{figure}[t]
\begin{center}
\psfig{figure=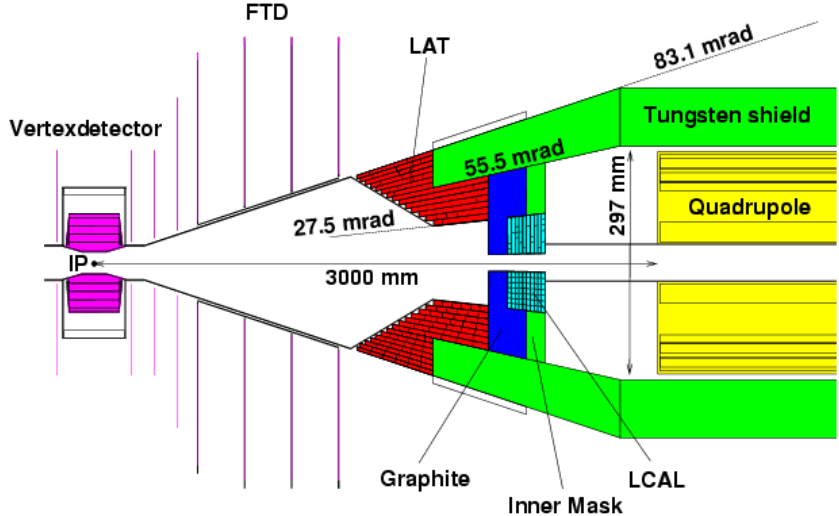,width=11truecm}
\end{center}
\vskip -5mm
\caption{\sl \label{fig:fwd_det}
 Forward region given in the TDR of TESLA indicating the various 
 vetoing components.}
\end{figure}

A detailed description of the forward equipment used in TESLA can be found 
in the TDR~\cite{TDR} and is shown in figure~\ref{fig:fwd_det}. 
The most relevant calorimeter 
for the present study is the LCAL, used as a beam monitor. In the TDR, this 
detector was situated at $2.6$\,m and started at $R=1.2$\,cm. 
In a recent upgrade of the final focus set up, the final quadrupoles 
have been moved downstream in such a way that the calorimeter is now at 
$3.7$\,m, at the same radius therefore covering an angle down to $3.2$\,mrad.

\par
\begin{figure}[htb]
\begin{center}
\psfig{figure=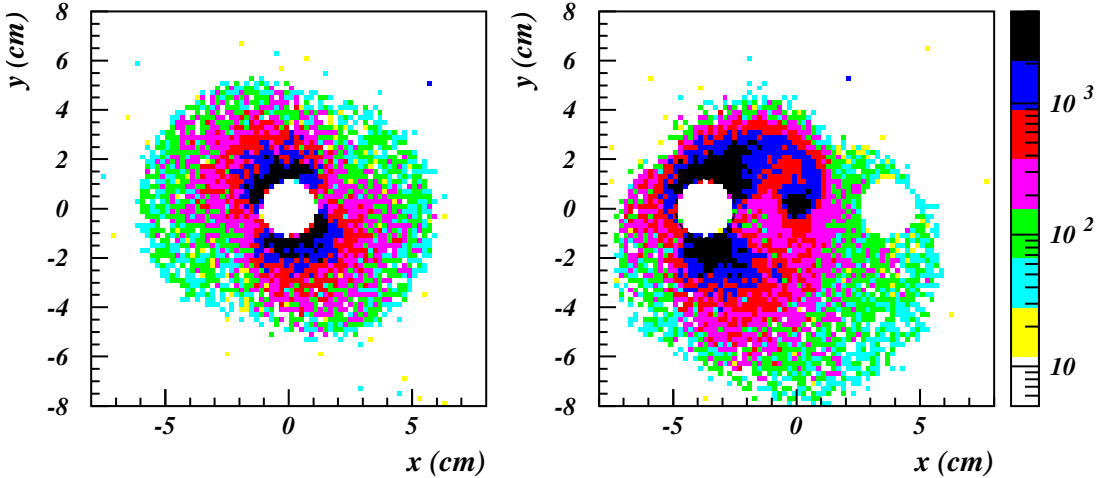,width=15truecm }
\end{center}
\vskip -15mm
\caption{\sl \label{fig:bg_density}
 Energy deposits in GeV from beamstrahlung in the TESLA configuration 
 without and with crossing angle. Location is at $3.7$\,m downstream of 
 the interaction point.}
\end{figure}

The energy density distribution of the background induced by beam-beam 
interactions is given in figure~\ref{fig:bg_density}.  
This distribution corresponds to 
one bunch crossing since in the TESLA configuration the calorimeter can be 
read before the next crossing.
The detection of energetic electrons is possible everywhere 
since these electrons can be recognized from the low energy electrons 
which dominate the background using the longitudinal energy profile of 
their shower. An optimal treatment of the vetoing procedure is still underway.
A detailed note on this problem should appear in the near future~\cite{veto}.
In the present work the LCAL was taken as an ideal veto but only for very 
energetic electrons (see section~\ref{sec:anastau}). 

\par
With a crossing angle, there are two blind regions of the vetoing device. 
One corresponds to the hole ($R=1.2$\,cm) needed for the entering beam, 
the other to the hole ($R=1.2$\,cm) needed for the exit of the disrupted beam. 
Both holes can also be surrounded with a vetoing device. 
Note also that due to the finite angle between the beam and 
the solenoidal field, the secondary electrons from the disrupted beam 
experience an azimuthally asymmetric curvature which also creates 
an asymmetric background distribution.

\section{Generators and tools used for this analysis}
The present analysis uses generators which were developed and tested at LEP2: 
SUSYGEN for the signal, BDKRC for the leptonic $\gamma\gamma$ background 
and Pythia including 
direct, VDM, anomalous and DIS sub-processes for the hadronic $\gamma\gamma$ 
background. The direct process stands for those 
interactions in which the bare photon interacts directly with 
its partner, the VDM for those where the photon fluctuates into a vector meson,
predominantly $\rho^0$, the anomalous (or generalized VDM) in which the
photon fluctuates into a $q\overline{q}$ pair of larger virtuality than 
in VDM process and the DIS corresponds to the deep inelastic scattering
process $\gamma^\ast q\rightarrow q$ with $q$ from the VDM and anomalous 
processes.

\begin{figure}[htb]
\begin{center}
\psfig{figure=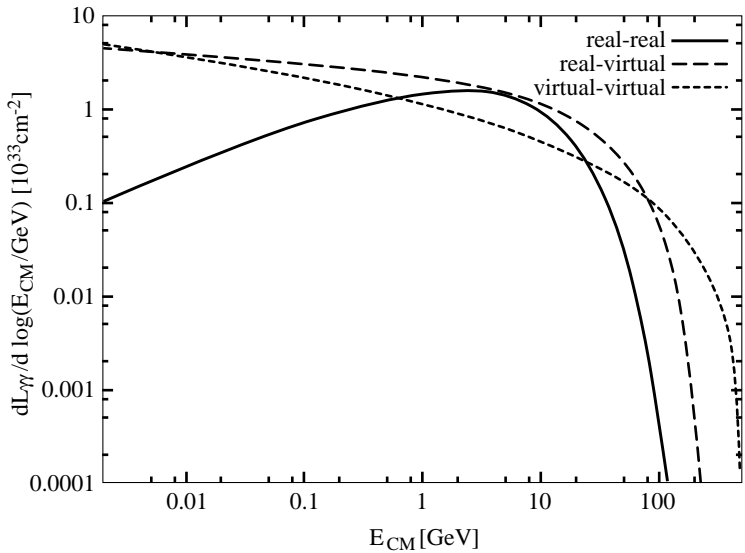,width=11truecm}
\end{center}
\vskip -10mm
\caption{\sl \label{fig:gg_sigma}
Differential luminosity corresponding to the various components of 
$\gamma\gamma$ physics~\cite{schulte}. The so-called `virtual-virtual' 
corresponds to the standard component induced by $e^+e^-$ interaction. 
The two others correspond to interactions of real photons produced through 
beamstrahlung and which interact either with similar photons from 
the opposite beam (real-real) or with electrons/positrons from the opposite 
beam (real-virtual). In the energy domain of interest for this analysis,
$\sim 10$\,GeV, one should multiply by 5 the rate due to the virtual-virtual 
contribution generated by our programs.}
\end{figure}

In addition to the previous processes with virtual photons one should 
take into account the processes due to real secondary photons induced 
by beamstrahlung. Figure~\ref{fig:gg_sigma} taken from \cite{schulte} 
shows that these spectra,
in the region of interest with photons between $1$ and $10$\,GeV, 
increase the standard background by a factor of order 5. 
One should however note that contrary to virtual photons, 
the real-real spectrum has no transverse momentum and is therefore 
much less dangerous. 

Given the very large background cross sections involved, a procedure of 
enrichment was applied. A fast simulation program, SGV~\cite{bergg}, 
was used since it provides a realistic and well tested simulation tool. 
This modelization was only used to define the acceptance of the detector 
for charged tracks and neutrals and to allow for secondary processes 
(e.g.\ conversion of photons) but no reconstruction effects were included 
(e.g.\ overlap of charged tracks and neutral deposits in the calorimeters). 

The various points defined in section~\ref{sec:dm_susy} were generated 
with emphasis on the most difficult ones which correspond to masses 
close to the energy threshold and with smallest mass difference 
between the slepton and the neutralino.  

\section{Analysis for the smuon channel}
\begin{figure}[htb]
\begin{center}
\psfig{figure=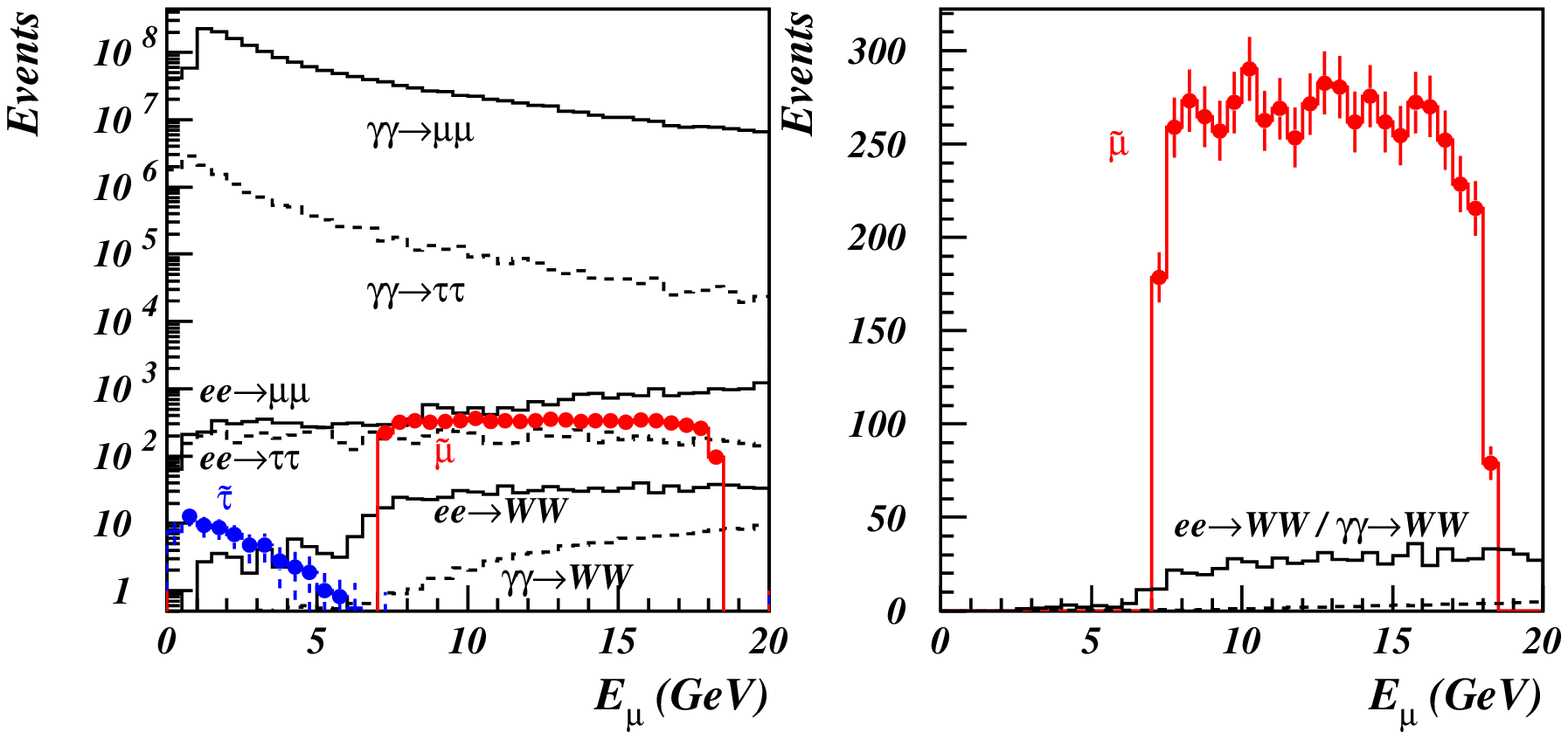,width=15truecm }
\end{center}
\vskip -10mm
\caption{\sl \label{fig:emu}
 The first plot shows the distribution of the muon energies of events 
 selected with two muons. The second plot shows the same distribution after 
 applying the cuts described in the smuon analysis.}
\end{figure}

This channel is much easier to treat than the stau channel since, 
as shown in figure~\ref{fig:emu}, the energy deposited by for smuons is 
much larger than for the lightest stau. 
This plot corresponds to the point D$^\prime$ 
which has been selected as one of the most challenging ones. 

The cuts used are the following (after applying the forward veto: 
two muons with transverse momentum greater than $2.5$\,GeV and with 
an azimuth difference below $160$ degrees and an overall missing transverse 
momentum larger than $5$\,GeV. These cuts retain $80\%$ of the signal.
The results are shown in figure~\ref{fig:emu} for an integrated luminosity of 
$500$\,fb$^{-1}$. Both energy edges emerge cleanly above a small background 
which allows a very precise determination of the slepton and LSP masses. 
The latter will of course be used as an input to predict the amount of 
dark matter in the universe.

\section{Analysis for the stau channel} \label{sec:anastau}
The working point chosen is D$^\prime$ (see Table~\ref{tab:scenarios}). 
This channel is primarily contaminated by $ee\rightarrow\tau\tau ee$  
which will produce topologies similar to the signal. 
The process ee$\rightarrow\mu\mu$ee can be  suppressed, with a very small 
loss of efficiency, by vetoing final states with two identified muons. 

In a small fraction of cases, however, such an event topology can occur
where one of the spectator $e^\pm$ is emitted at relatively large angle and 
is misidentified as $\tau\rightarrow e\nu_e\nu_\tau$ decay while one of 
the real $\mu$ or $\tau$ goes to lower angles down to $20\,$mrad and 
cannot be detected by any detector unless the LAT (figure~\ref{fig:fwd_det}) 
has the capability of detecting a $\mu$ or $\pi$. For the present analysis, 
these background events
are rejected by excluding $eX$ topology. However, given the low analysis 
efficiency (table~\ref{tab:results}), it would be highly desirable to 
consider the possibility of providing an efficient
$\mu/\pi$ identification in the forward instrumentation.

The $WW$ background can also give a significant contribution to the
stau analysis, via the $\tau\nu$ decay of the $W.$ 
In practice the momentum and angular distributions of these $\tau$ particles 
behave differently from those of the signal which give very soft and
isotropically distributed particles. 
Na\"{\i}ve selections leave only a few events and 
with a likelihood method (not implemented in the present analysis) 
one should essentially end up with a negligible background contribution
(this would not of course be true for a large mass difference scenario). 
\begin{figure}[htb]
\begin{center}
\psfig{figure=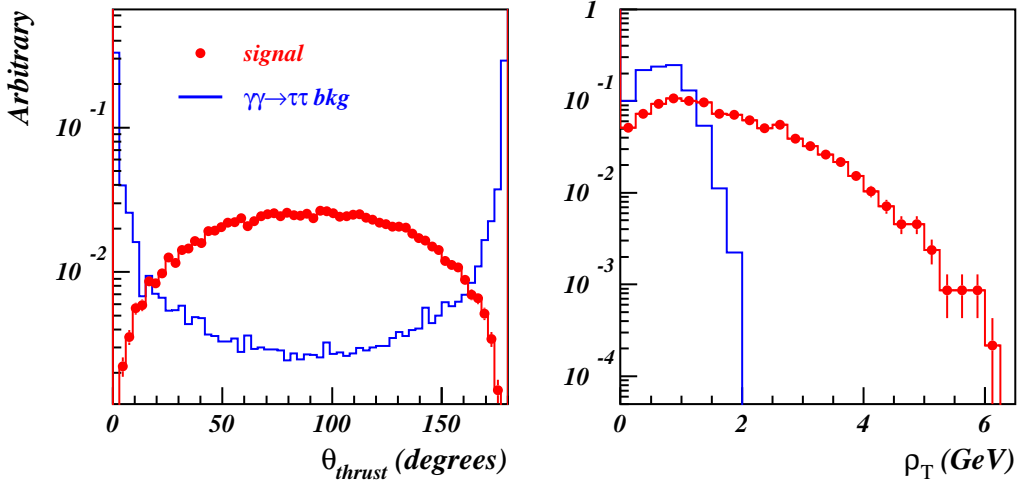,width=15truecm }
\end{center}
\vskip -10mm
\caption{\sl \label{fig:tau_shape}
 The first plot shows the polar angle distribution of the thrust axis 
 for the signal and for the background. Statistical fluctuations on background
 reflect the weighting procedure. The second plot shows the distribution 
 for the variable $\rho_T$ defined in the text and is obtained by adding 
 all other cuts except the bi-dimensional cut shown in figure~\ref{fig:cuts}.}
\end{figure}

A thorough investigation of the various hadronic sub-processes has also 
been carried out \cite{berggh}. The dominant background comes from 
the process $ee\rightarrow c\overline{c}ee$, where one of the charm quarks 
decays semi-leptonically.   

\subsection{The leptonic background} \label{sec:lept_bg}
The signal, as can be seen in figure~\ref{fig:emu}, leaves a very small 
amount of visible energy as compared to the smuon analysis.
The main selection relies on the fact that in the transverse plane 
the background $\tau$ leptons are back to back while they are uncorrelated 
in azimuth in the case of the signal. The crossing angle effect basically 
does not modify appreciably this situation when the final state electrons 
are properly vetoed. 
Using this feature, one can reconstruct the common direction of 
the $\tau$ particles by defining a thrust axis in the transverse plane. 
Then one computes $\rho_T$, the sum of the modules of the transverse 
momenta of individual particles with respect to this axis.
The distribution of this quantity is displayed in figure~\ref{fig:tau_shape}, 
showing a clear separation between the $\tilde{\tau}$ signal and 
the $ee\rightarrow \tau\tau ee$ background. 

\par
The list of cuts given below is meant to remove both the leptonic and 
the hadronic backgrounds (e.g. veto on K$^0_L$ and neutrons): 
\begin{enumerate}
\item veto on photons and electrons with a transverse momentum above 
 $0.8$\,GeV and within $15$\,degrees with respect to the beam axis 
\item request a $\tau^+\tau^-$ topology: 
  \begin{itemize}
   \item $1-1$, $1-3$, or $3-3$ prongs except for the $eX$ and $\mu\mu$ 
     topologies
   \item if $3-$prong, then no additional visible neutral particles
   \item if there is a muon then only $1-$prong and 
     no additional visible neutral particles
   \item no $K^0_S$, $K^0_L$ and neutron in the event
   \item visible mass of the tau below $2\,$GeV
   \item correct charge in each hemisphere and total charge conserved
  \end{itemize}
\item polar angle of the thrust axis ($\theta_{\rm thrust}$)
   between $15-165$ degrees and
   acoplanarity angle ($\phi_{Acoplanarity}$) below $145$ degrees
\item maximum particle momentum below $15$\,GeV and
   the missing transverse momentum ($P_{T\,miss}$) of the event greater than 
   $2.5$\,GeV
\item combined rejection on $\phi_{Acoplanarity}$ and $\rho_T$ as indicated 
 in figure~\ref{fig:cuts}.
\end{enumerate}
In the case of a zero crossing angle and using the vetoing procedure defined 
in section~\ref{sec:fwd}, one remains with one tau background event 
with a weight of slightly below one and 
an efficiency of $6.3\pm 0.2\%$ for $\sqrt{s}=500$\,GeV 
(table~\ref{tab:effevt}).
This result does not depend significantly on the beam energy as was shown 
by generating the same sample at $442$\,GeV (table~\ref{tab:results}).
\begin{table}[htb]
\caption{\sl \label{tab:effevt}
 The efficiency, the signal and dominant background events in
 the head-on case for working point D$^\prime$ at $\sqrt{s}=500$\,GeV.}
\begin{center}
\begin{tabular}{|c|c|c|c|}
\hline 
 Efficiency ($\%$) & $N(\tilde{\tau}\rightarrow \tau\chi)$ & 
  $N(ee\rightarrow \tau\tau ee)$ & $N(ee\rightarrow q\overline{q}ee)$ with $q=c,b$ \\
\hline
 $6.3\pm 0.2$ & $316\pm9$ & $1.0\pm 1.0$ & $1.0\pm 1.0$ \\
\hline 
\end{tabular}
\end{center}
\end{table}

\begin{figure}[htb]
\begin{center}
\psfig{figure=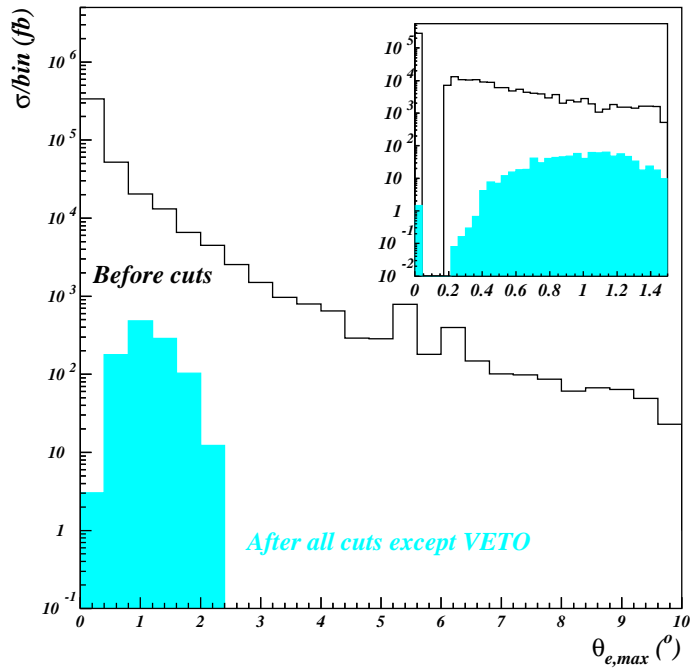,width=10truecm }
\end{center}
\vskip -7mm
\caption{\sl \label{fig:e_theta}
Angular distribution of the spectator electrons from 
$ee\rightarrow\tau\tau ee$ expressed in fb/bin. The window shows the same 
distribution in the very forward region. In the inset, the peak at zero 
corresponds to those events where both electron and positron spectators
stay in the beam-pipe. The light shaded distribution 
corresponds to the distribution obtained after all the selections described 
in the text, with the exception of the forward veto.}
\end{figure}

The role of the efficiency veto is crucial as shown in 
figure~\ref{fig:e_theta}. 
For angles above $0.5$ degrees, one needs a rejection to better than 
a thousand while maintaining a fake rate at a reasonable level. 
Since the selection is applied not on energy but on transverse momentum 
one sees that for the smallest angles this veto is only requested for 
very energetic electrons, therefore easily recognized from the main background.
As mentioned in section~\ref{sec:fwd}, there are ongoing studies to 
demonstrate the feasibility of this device.

\begin{figure}[htb]
\begin{center}
\psfig{figure=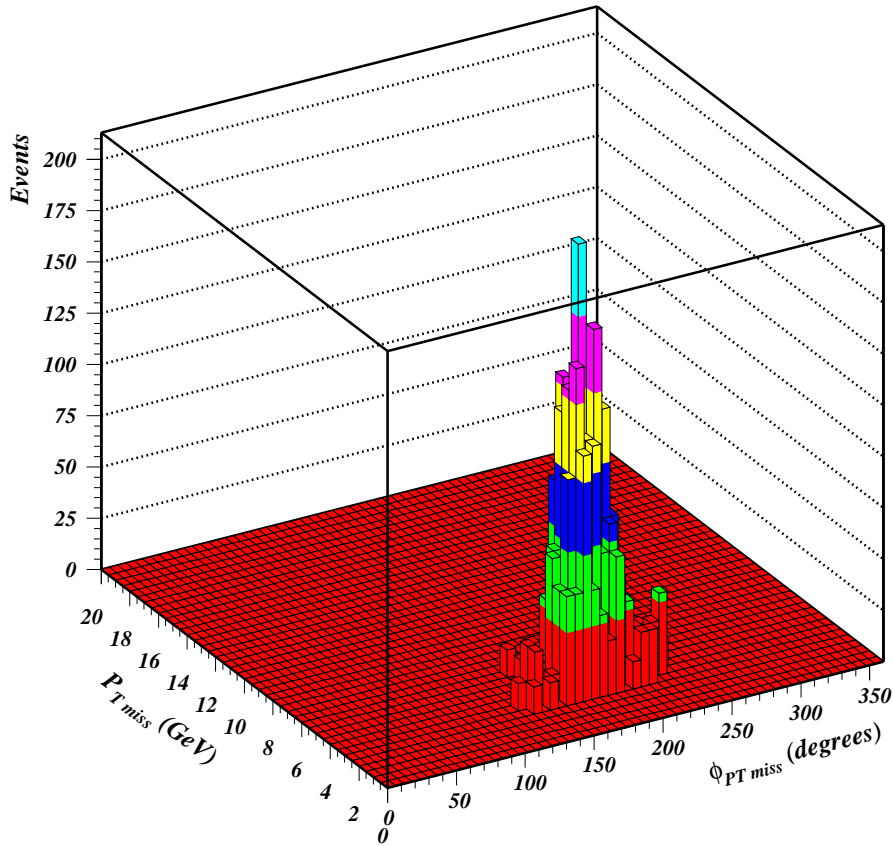,width=13truecm }
\end{center}
\vskip -10mm
\caption{\sl \label{fig:bg_xangle}
 Remaining $\gamma\gamma$ background events based on 
 $ee\rightarrow \tau\tau ee$ with all cuts as for 
 the head-on analysis. $P_{T\,miss}$ is the resulting transverse momentum of 
 the visible particles, while $\phi_{PT\,miss}$ is the azimuthal orientation 
 of this momentum.}
\end{figure}

\begin{figure}[htb]
\begin{center}
\psfig{figure= 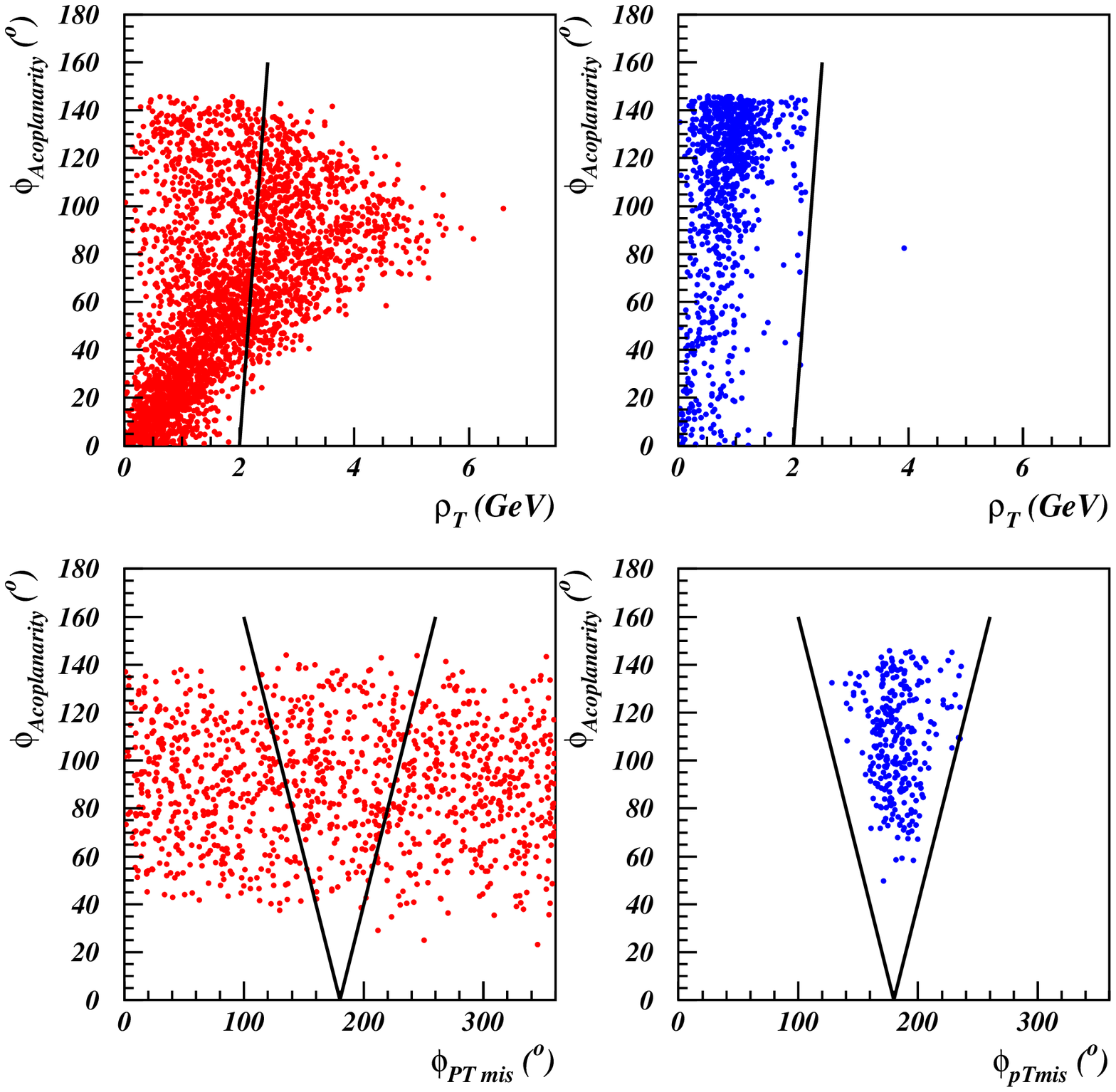,width=13truecm}
\end{center}
\vskip -10mm
\caption{\sl \label{fig:cuts}
 Selections on the acoplanarity angle and on the $\rho_T$ variables 
 described in the text are displayed on top for the signal (in red, left plot) 
 and on the background (in blue, right plot). Below are illustrated 
 the variables used for the additional cuts needed with a crossing angle.}
\end{figure}

In the case of a crossing angle, and without modifying above selection, 
one finds $4000\pm 225$ events for the $ee\rightarrow\tau\tau ee$ background 
with a luminosity of $500$\,fb$^{-1}$. The large majority of these events 
are easily identified as shown in the plot of figure~\ref{fig:bg_xangle}. 
They are essentially due to events for which the final state 
electron/positron ends up in the `wrong hole' i.e.\ in the entrance hole 
for the opposite beam. In such cases the $\gamma\gamma$ process will have 
an unbalanced transverse momentum of about 5 GeV pointing in the horizontal 
plane. This background can be almost entirely eliminated by a combined cut on 
$\phi_{Acoplanarity}$ and on the angle ($\phi_{PT\,miss}$) of the missing 
transverse momentum vector plane, as indicated in figure~\ref{fig:cuts}. 
There is then a relative reduction in efficiency of $25\%$. 

How often does an electron end up in the wrong hole? 
This fraction depends on the size of the hole but is much larger than
one would estimate on the basis of the solid angle. Typically one finds that 
the probability is $10^{-3}$, which is certainly not negligible given 
the rates. 

Should one increase the energy to improve on the efficiency? 
Yes but in practice there could be no choice if there is a heavy 
$\tilde{\tau_1}$. Moreover, as pointed out in the next section, 
when the energy is increased one looses rapidly the sensitivity in determining
the tau slepton mass.

\subsection{The hadronic background}
\begin{figure}[htb]
\begin{center}
\psfig{figure=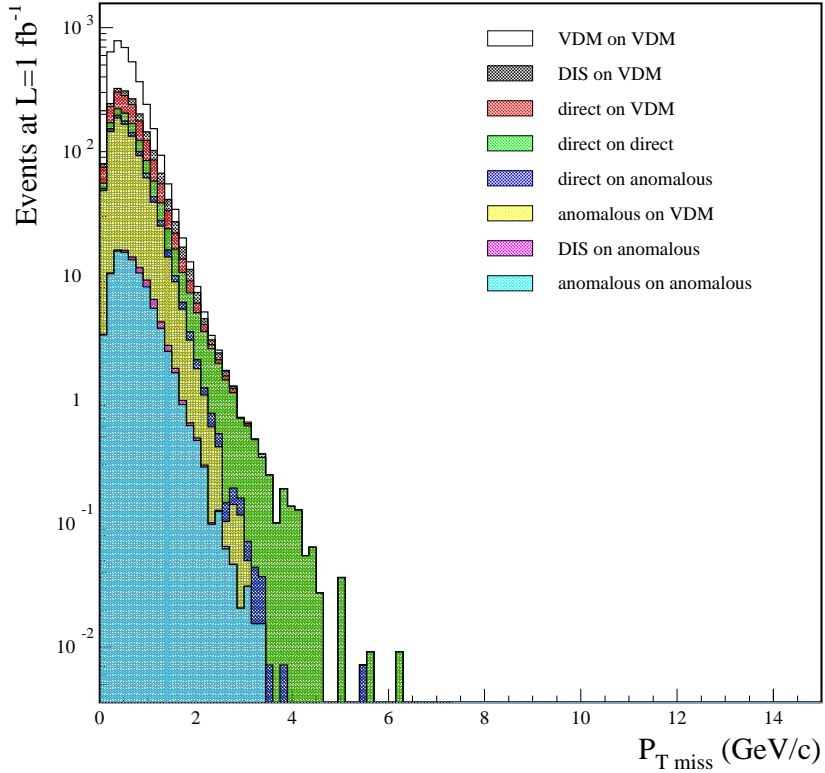,width=12truecm}
\end{center}
\vskip -7mm
\caption{\sl \label{fig:hadrons}
 Comparison of the different hadronic sub-processes indicating that 
 the direct on direct process (green)
 gives the largest missing transverse momenta.}
\end{figure}

Hadronic sub-processes have an even larger cross sections but most have 
a topology distinct from the signal and, in particular do not produce 
appreciable missing transverse momenta. This feature appears clearly 
in figure~\ref{fig:hadrons} where the various hadronic sub-processes 
have been simulated~\cite{berggh} using Pythia.
The major contribution comes from $ee\rightarrow c\overline{c}ee$ in 
direct on direct interactions, 
where the missing transverse momentum originates from the semi-leptonic 
decay of a charm particle. 
With the selections described in the preceding section one hadronic 
background event was left with an event weight of again slightly below one
(table~\ref{tab:effevt}).
\subsection{Mini-jets}
The influence of mini-jets is clearly machine dependent.  
In the case of TESLA, there is a $25\,\%$ probability to have a mini-jet 
with energy greater than $5$\,GeV per beam crossing. 
One therefore expects that in a few \% of cases there could be 
two mini-jets superimposed. Such topologies are clearly not generated 
by standard codes and, for the present study, one can already attempt to crudely evaluate
this contribution. \par
To do this, 
one generates mini-jet events and evaluates the probability $p_\tau$ that
they produce a topology consistent with a single tau. One finds
$p_\tau\sim5\cdot 10^{-4}$. Then one combines two such events occurring 
in the same crossing, with a probability of $3\,\%$ in the case of TESLA. 
Using the micro-vertex information, one eliminates mini-jets with
inconsistent vertex. This keeps about $1\,\%$ of the events ($3$ times more
in the case of a warm machine with a shorter bunch length). 
One applies the series of cuts defined previously which keeps about
$10\,\%$ of these events. Finally one eliminates one half of the remaining
events on the basis of charge conservation.

Taking into account the number of crossings corresponding to
$500$\,fb$^{-1}$, one finds that this background gives about one event and 
therefore does not modify our overall conclusions.
For the NLC technology the result depends on the number of bunches to
be integrated. With a number of mini-jet/crossing $2.5$ times smaller
but with a bunch length $3$ times shorter, NLC would give $\sim 0.5 N_b$ 
times the previous result, where $N_b$ is the number of
bunch within the time resolution which in our opinion should be at
least $3$.
\subsection{Discussions}
There are three caveats to the present studies:
\begin{itemize}
 \item an ideal reconstruction and a perfect veto efficiency down to 
  $3.2$\,mrad were assumed with a transverse momentum threshold at $0.8$\,GeV,
 \item only the dominant $\gamma\gamma$ background processes $\tau\tau$ 
  and $q\overline{q}$ (with $q=c$ and $b$) were studied with 
  sufficient statistics,
 \item no effect of the machine backgrounds\footnote{The machine background
  originating from beamstrahlung photons is, however, machine
  dependent (figure~\ref{fig:gg_sigma}).
  The hadronic background is also subject to uncertainties on the cross
  section for hadron production in photon-photon collisions.} 
  were taken into account, 
  in particular the overlapping mini-jets should also be considered 
  since they could affect the reconstruction of the standard background and 
  generate some tails in the distributions shown in 
  figures~\ref{fig:tau_shape} and \ref{fig:bg_xangle}. 
\end{itemize}

With these caveats in mind, one concludes that in a LC even in this 
difficult case, clean $\tilde{\tau}$ samples with no significant background
can be obtained and the mass of the lightest tau slepton can thus be 
determined (section~\ref{sec:stau}).
For the co-annihilation scenario, 
this result allows a model independent and precise prediction of the DM 
content of the universe. 
 
\section{Prediction of the DM content of the universe}
In the co-annihilation scenario, the DM content of the universe depends 
primarily on two quantities:
\begin{itemize}
 \item the mass of the LSP which can be determined, to a high accuracy, 
  either using the chargino/neutralino or the smuon/selectron channel
\item the mass of the lightest sleptons, and in particular the mass of 
  the lightest stau which is determined with less precision.
\end{itemize}
These statements are general and do not assume any particular SUSY model 
like mSUGRA. In contrast, as was proposed in \cite{poles} for an LHC analysis,
one assumed mSUGRA and extracted the scalar and gaugino masses from 
the cleanest observables (the smuon channel in the case of the LC). 
This approach, necessary at LHC, is clearly model dependent but, as shown 
in this paper, can be avoided in most cases in the LC environment.

\subsection{Measurement of the stau mass using the threshold method}
\label{sec:stau}
\begin{figure}[htb]
\psfig{figure=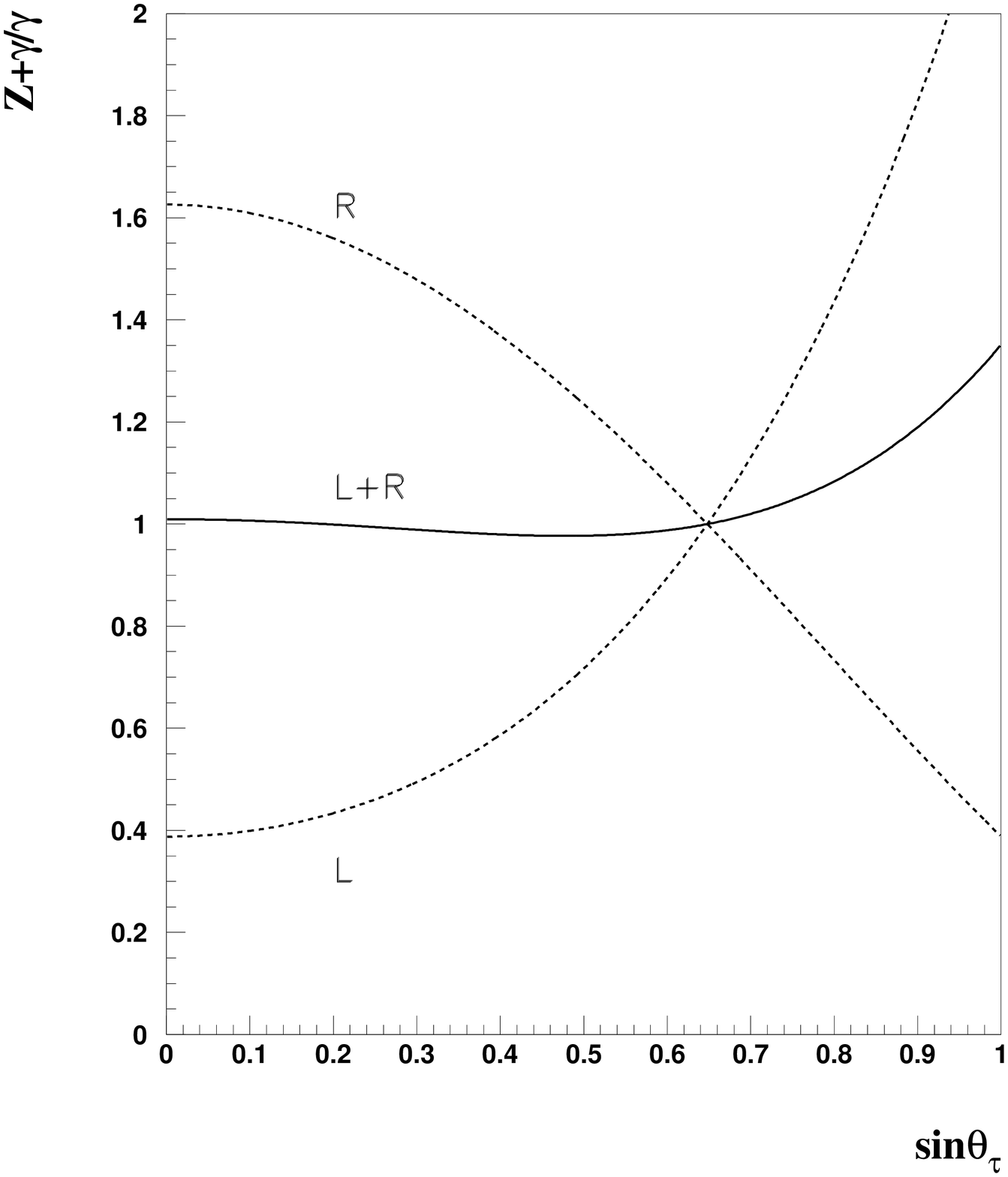,width=8truecm}
\psfig{figure=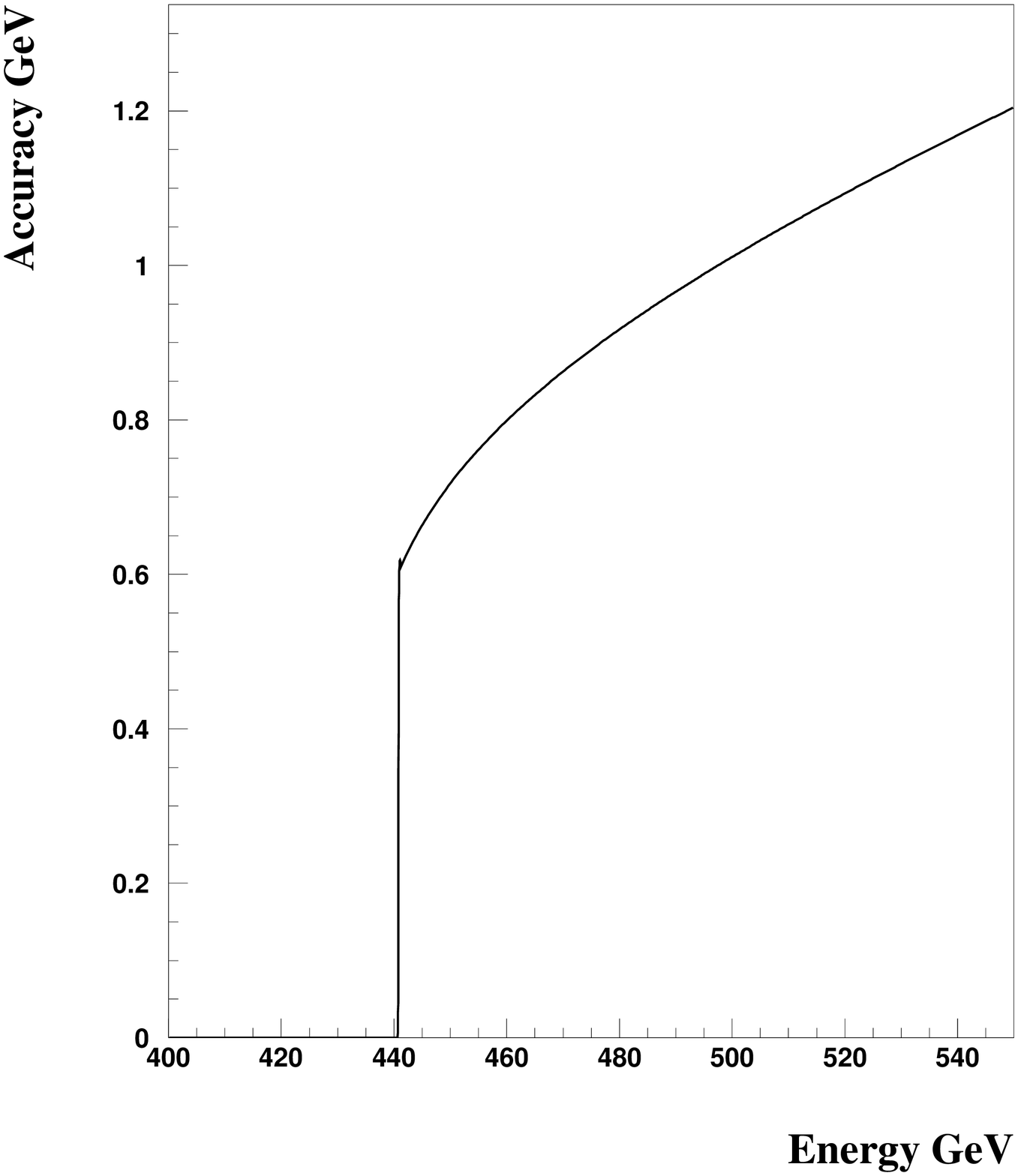,width=8truecm}
\caption{\sl \label{fig:stau}
The first plot gives the ratio between the stau cross sections, 
including the $Z$ and the photon exchange and including only 
the photon exchange, versus the sine of the mixing angle in three cases: 
unpolarized beam (full line), right-handed electrons 
(dashed curved labeled $R$) and left-handed electrons (labeled $L$). 
The second plot gives the precision on the stau mass versus 
the center of mass energy for the working point D$^\prime$. 
The cut near threshold reflects our choice to collect at least $10$ events 
with $500$\,fb$^{-1}$. }
\end{figure}

To extract the $\tilde{\tau}_1$ mass with minimum luminosity, 
the method consists in measuring the cross section at one energy and deduce 
the mass from the value of $\beta$ since, at the Born level, 
this cross section depends on $\beta^3=(1-4m^2/s)^{3/2}$, where $m$ 
stands for the stau mass. One can also assume, as shown in 
figure~\ref{fig:stau}, 
that the unpolarized cross section has very little dependence 
on the stau mixing angle which is true for reasonable values of 
the mixing angle, i.e.\ for a mixing angle below $\pi/4$. 
Above this value the lightest stau would be dominantly left-handed 
but this behavior is revealed with a polarized electron beam as shown 
in figure~\ref{fig:stau}. This effect is so strong that with a sample of 
$50$ events one could exclude such a possibility at the 6 s.d.\ level.

At which energy should one operate to achieve the best accuracy? 
One finds (see Appendix) that, without background and for a given integrated
luminosity $L$ (in fb$^{-1}$), the best accuracy is obtained very near 
threshold. 
The optimum is set by the requirement to observe a significant number of 
events, say at least $N>10$ events. Then one finds that the optimum 
in energy is given by the condition  $\beta^3=N/(LA\epsilon)$ while 
the corresponding relative accuracy on the mass is given by:
$$ \delta=\frac{(N)^{1/6}}{3(LA\epsilon)^{2/3}}\,.$$  
where $\epsilon$ is the efficiency on the stau channel and $A$ is a constant 
approximately equal to $100$\,fb at $\sqrt{s}=500$\,GeV and varies like 
$1/s$. The above formula shows that there is a very slow dependence of 
the precision on the choice of $N$. If one requires $50$ events instead of 
$10$ to cross check the mixing hypothesis with polarized electrons, 
the same precision could be achieved by simply increasing the luminosity by 
$50\%$. 

For point D$^\prime$ this gives $\beta\simeq 0.19$ corresponding to a 
$\sim 0.5$\,GeV error on the stau mass for $500$\,fb$^{-1}$ and 
an optimum $\sqrt{s}$ at $\sim 442$\,GeV. 
The gain in luminosity with this choice is appreciable 
as can be seen in figure~\ref{fig:stau}. 
Without optimization the error would have been $1.2$\,GeV.

Note that to achieve an optimum, one should have an a priori estimate of 
the stau mass and of the efficiency. 
This estimate can be guided by the results on the first two generation 
sleptons which also provide a determination of the LSP mass. 
For large value of tan$\beta$, a precise estimate of the stau-LSP mass 
difference and therefore of the efficiency, could be delicate since, 
as can be seen in table~\ref{tab:scenarios}, there can be large differences 
between the stau mass and the right-handed ordinary slepton masses. 
As previously mentioned in section~\ref{sec:dm_susy}, 
one could miss the smuon information 
in a scenario for which the first two generations are very heavy. 
In such cases, it is still possible to work out a strategy by 
first operating at the maximum energy to observe the stau signal and 
get a first estimate of the stau mass from the cross section 
which would allow to define the adequate center of mass energy 
for an optimal precision. For the LSP mass one should then use 
the chargino/neutralino reactions. 

The simple formulae used for this discussion did not take into account 
ISR effects which are important for the stau cross section 
near threshold. It can be shown (see Appendix) that the effects on
the stau cross section can be represented by a correction factor 
$0.86\beta^{2x}$ where $x$ is 
the virtual radiator given by $x=2\alpha/\pi[\log(s/m^2_e)-1]$. 
As an example, this correction applied on point D$^\prime$ gives 
a corrective factor of $0.53$ for the optimized value $\beta\simeq 0.19$. 

Results are summarized in table~\ref{tab:results} for point D$^\prime$ as well
as for other relevant points assuming that $500$\,fb$^{-1}$ 
has been taken at the optimized energies which are indicated.
These results are obtained assuming that there is negligible background. 
\begin{table}[htb]
\caption{\sl \label{tab:results}
 Error on the mass difference between the stau and the LSP with 
 $500$\,fb$^{-1}$ luminosity (and an optimal choice of energy) for the TESLA 
 assumptions. Effect on the relative uncertainty on $\Omega_{DM}h^2$.}
\begin{center}
\begin{tabular}{|c|c|c|c|c|c|}
\hline 
 Model & A$^\prime$ & C$^\prime$ & D$^\prime$ & G$^\prime$& J$^\prime$ \\
\hline
\hline Optimal $\sqrt{s}$ GeV & $505$ & $337$ & $442$ & $316$ & $700$ \\
\hline Efficiency in $\%$ & $10.4$ & $14.3$ & $5.7$ & $14.4$ & $<1.0$ \\
\hline Error on mass GeV & $0.487$ & $0.165$ & $0.541$ & $0.132$ & $>1.0$ \\
\hline Error on $\Omega_{DM}h^2$ in $\%$ & $3.4$ & $1.8$ & $6.9$ & $1.6$ & 
 $>14$ \\
\hline 
\end{tabular}
\end{center}
\end{table}

In this discussion, one has ignored the energy spread of the effective 
luminosity which is machine dependent. This issue is common to various 
studies performed at the future LC (e.g.\ top threshold, SUSY thresholds) 
and precise methods are being developed to determine this differential 
luminosity based on $e^+e^-$ scattering. 
The analysis described here is not more demanding in this respect than those
developed e.g.\ for the top threshold. Thus the energy spread effects
will not be investigated any further.

\subsection{Determination of the SUSY DM component}
The program Micromegas~\cite{micro} has been used to compute the uncertainty 
on the DM density due to the SUSY mass error measurements. 
This program operates without any assumption, in particular it does not 
rely on the mSUGRA scheme.

Results are listed in table~\ref{tab:results}. It shows, as expected, 
that $\Omega_{DM}h^2$ depends primarily on the precision on the stau and 
LSP masses (through the Boltzmann law, as explained in the introduction).  
The present analysis, 
developed for the D$^\prime$ solution, gives satisfactory results 
except for point J$^\prime$ which is almost beyond detectability. 
It is however fair to say that no effort was invested to adapt this analysis 
for the J$^\prime$ case. Points H$^\prime$ and M$^\prime$ are omitted since 
they do not pass the WMAP constraints within Micromegas. Points B$^\prime$, 
E$^\prime$, F$^\prime$, I$^\prime$, K$^\prime$ and L$^\prime$ are irrelevant 
since they have mass differences above 10 GeV and are easy to measure. 
One should notice that the point D$^\prime$ itself, which corresponds to 
a negative value of $\mu$, is marginally compatible with the $g-2$ result. 

An alternative method, as previously noted, would be to work at the maximal 
energy of the collider and collect a large sample of events to analyze 
the high energy spectrum to estimate the mass of the stau. 
This could be done, with a 500 GeV collider, for point C$^\prime$ and 
G$^\prime$. It is however worth noting that the precision achieved with 
the threshold method is quite challenging and a comparison of the two methods 
should include systematic uncertainties. 


\subsection{A focus type solution} \label{sec:focus}
Table~\ref{tab:focus} gives the parameters of a `focus' type solution taken 
from reference \cite{baer}. This solution predicts a light neutralino and 
the lightest chargino about $30$\,GeV heavier. One can therefore expect 
a clean signal and assume the mass accuracies given in \cite{TDR-physics}.
\begin{table}[htb]
\caption{\sl \label{tab:focus}
 Accuracies expected in the `focus' scenario taken from \cite{baer}.}
\begin{center}
\small
\begin{tabular}{|c|c|c|c|c|c|c|c|c|}
\hline 
 Parameters  & $M1/2$ & $m0$ & $\tan\beta$  & $\mu$ & $m_{\chi_1}$ & 
 $m_{\chi_2}$ & $m_{\chi^{\pm}_1}$ & $m_{\chi^{\pm}_2}$
 \\
\hline
\hline Values ($m, \mu$ in GeV) & $300$ & $2500$ & $30$ & $121.6$ & $85.6$ & $135$ &
 $113.1$ & $274.8$ \\
\hline Accuracies ($m, \mu$ in GeV) & $-$ & $-$ & $+25-11$ & $1.4$ & $0.1$ & $0.3$ & $0.04$ &
 $0.25$ \\
\hline Error on $\Omega_{\rm DM}h^2$ in $\%$ & $-$ & $-$ & $+8.6-5.9$ & 
 $+2.9-2.1$ & $-$ & $-$ & $-$ & $-$ \\     
\hline 
\end{tabular}
\end{center}
\end{table}
One should be able to achieve threshold scans for the charginos and 
to access the mass difference through the dilepton mass distribution 
given by the $\chi_2\chi_1$ channel. One can then use the cross section 
measurement and the polarization asymmetry in the $\chi^+_1\chi^-_1$ channel
to measure the parameter $\mu$ which governs the estimate of DM. 
This has been done assuming the accuracies given in \cite{TDR-physics} and 
the formulae taken from \cite{chargino}.\par

With these very precise chargino/neutralino masses, the main source of 
uncertainty is due to the cross section measurement and the main error 
on $\Omega_{\rm DM}h^2$ is due to $\tan\beta$ since there is  
a poor sensitivity for high values of $\tan\beta$. 
There is no other practical mean to determine this quantity, given that 
the sfermions and the heavy Higgses are inaccessible in the focus scenario. 
In spite of this, the precision achieved on $\Omega_{\rm DM}h^2$ would be 
largely sufficient to demonstrate a contradiction
with the WMAP result and therefore imply that there are other sources of DM.

One therefore concludes that even if the observation of the chargino/neutralino
sector is easy at a LC (and perhaps only achievable at a LC) in a focus 
scenario, the accuracy on 
the indirect determination of the DM content of the universe requires 
the highest possible accuracies. It is however 
fair to say that much more work is still needed to cover this scenario.

\section{Conclusions}
This analysis has shown that the detection and the mass measurement of 
the tau slepton, potentially important in view of the cosmological 
implications, is challenging in the so-called `co-annihilation' scenario. 
A forward veto to remove the $\gamma\gamma$ background down to very small 
angles is essential to reach an almost background free result, adequate to 
achieve the accuracy implied by the post-WMAP generation in a model 
independent analysis. 

\par
\vspace{2mm}
From the present analysis, which includes the relevant backgrounds but 
not yet a fully realistic modelization of the detector response, 
one can already state that: 
\begin{enumerate}
\item in the zero angle crossing situation, only possible in the TESLA scheme,
 the detection of the stau particles can be done with almost negligible 
 background,
\item for these solutions, the requirement to match the Planck era precision 
 demands luminosities of at least $500$\,fb$^{-1}$
 even with an optimized strategy of scanning. This further strengthens 
 the need for a collider able to deliver the maximum luminosity, 
 not necessarily at the maximum energy ,
\item in the TESLA case, with a half crossing angle of $10$\,mrad, 
 there is only a $25\%$ degradation of the efficiency which therefore 
 leaves open this possibility,
\item in the NLC case, the same conclusion could be reached provided that 
 there is no degradation due to pile-up of several bunches in the forward 
 veto (this may requires some R$\&$D for a very fast calorimeter). 
\end{enumerate} 
\vspace{6mm}

\section*{Acknowledgements}
Very useful contributions to the forward veto issue have been provided by 
K.~Buesser, K.~Moenig and A.~Stahl. They are gratefully acknowledged. 
This study has also benefited from interesting suggestions and criticisms 
on the physics analysis aspects from G.~B\'elanger, U.~Martyn, 
G.A.~Moortgat-Pick and M.~Peskin.

\newpage
\vspace{20mm}\begin{center}
{\Large \bf APPENDIX }\\
\end{center}
\vspace{8mm}
1/ The ratio R=$\sigma/\sigma_{\gamma pointlike}$ for a stau is given by:
$$ R=0.215 \beta^3[1+(L+R)C_{\tilde{\tau}}/(1-K_Z)+(R^2+L^2)C^2_{\tilde{\tau}}/2(1-K_Z)^2]$$
with $K_Z=M^2_Z/s$, $s^2_W\sim 0.21$, $L=(-0.5+s^2_W)/s_Wc_W$, 
$R=s^2_W/s_Wc_W$and $C_{\tilde{\tau}}=(-0.5s^2_{\tilde{\tau}}+s^2_W)/c_Ws_W$ 
where s$_{\tilde{\tau}}$ gives the mixing angle of the lightest stau 
(for $s_{\tilde{\tau}}$=0 one has a pure $\tilde{\tau_R}$). 
From above formula one deduces easily that for the weak or non mixing case 
one has:
$$ R=0.215 \beta^3[1-0.1/(1-K_Z)+0.1/(1-K_Z)^2]$$
which clearly shows that there is cancellation of the $Z$ contribution 
($K_Z$ can be neglected at high energy). For maximum mixing, 
$s^2_{\tilde{\tau}}=0.5$, one has $C_{\tilde{\tau}}\sim0$, which also gives 
a negligible $Z$ contribution. This behavior is seen in figure~\ref{fig:stau}. 

With right-handed electrons, one has
$$R_R\sim0.215 \beta^3[1+1.03C_{\tilde{\tau}}/(1-K_Z)+0.27C^2_{\tilde{\tau}}/(1-K_Z)^2]\,,$$
and with left-handed electrons:
$$R_L=0.215 \beta^3[1-1.42C_{\tilde{\tau}}/(1-K_Z)+0.51C^2_{\tilde{\tau}}/(1-K_Z)^2]\,.$$ 
These two components behave very differently as can be seen in 
figure~\ref{fig:stau}. 
One can therefore easily distinguish between a standard scenario 
where the lightest stau is dominantly right-handed for which there is 
no need to correct for mixing effects for the unpolarized cross section 
(or equivalently the average of the cross sections obtained with 
the two electron chiralities) and the opposite scenario for which 
a correction is needed if the $LR$ asymmetry is of opposite sign and 
differs significantly from zero.
 
\vskip 0.7 cm
2/ The ISR effect can be treated simply using a soft photon approximation 
re-summed at all orders. One has the integral:
$$  x\int_0^{th}u^{x-1}(\beta/\beta^0)^3du  $$
with $u=k/E$ where $k$ is the photon energy, $E$ the beam energy and where 
the integral is taken from 0 to the stau threshold.
$x$ is the effective radiator:
$$ x=2\alpha/\pi[\log(s/m_e^2)-1]$$ 
with $x\sim$ 0.124 at $\sqrt{s}$=500 GeV. \par 
The effect is to replace $\beta^3$ in the stau cross section 
(p-wave dependence) by:
$$ x\Gamma(x)\Gamma(5/2)\beta^{3+2x}/\Gamma(x+5/2) $$
which is $\sim 0.86\beta^{3+2x}$ at $\sqrt{s}$=500 GeV. 

\par
\vskip 0.7 cm 
3/ To discuss in a simple way energy optimization for the stau mass 
determination, ISR corrections will be ignored.
Given the efficiency $\epsilon$, the luminosity L, and the cross section 
$\sigma=A\beta^3$, one can write the number of events produced as 
$N=LA \beta^3\epsilon$. The cross section depends on the stau mass 
through $\beta$ and therefore one can simply translate the statistical error 
on $N$, $\sqrt{N}$, into a relative error on the mass $\delta$ through 
the formula:
$$ \sqrt{N}=\sqrt{LA\beta^3\epsilon}=\frac{12m^2}{s}LA\epsilon\beta\delta\,.$$
One easily deduces that:
 $$\delta=\frac{s}{12m^2}\sqrt{\frac{\beta}{LA\epsilon}}\,.$$
The dependence on $s$ and $\beta$ shows that, with all other parameters fixed, 
the precision improves when one operates near threshold. 
The limit of this optimum is set by the requirement to observe 
a significant number of events, say at least $N>10$ events. 
Then one easily finds that the optimum in energy is given by the condition  
$\beta^3=N/(LA\epsilon)$, while, from above formula, the accuracy is:
$$ \delta=\frac{(N)^{1/6}}{3(LA\epsilon)^{2/3}}\,,$$
where the approximate relation $s=4m^2$ near threshold 
has been applied.
This formula illustrates two important features of the optimum method: 
\begin{enumerate}
 \item the relative precision on the mass has a weak dependence on the choice 
  of $N$ (going from $10$ to $20$ degrades $\delta$ by $12\%$)
 \item the improvement in accuracy scales like $L^{2/3}$ and not like 
  $L^{1/2}$ in the absence of an optimization.
\end{enumerate}
The intuitive reason for the latter is that, with an increased luminosity 
and a fixed number of events, one can work at lower energy and therefore 
increase the sensitivity. 

For point D$^\prime$ this gives $\beta\simeq 0.19$ and a $\sim 0.5$\,GeV error
on the stau mass for $500$\,fb$^{-1}$ for an optimum energy $\sim 442$\,GeV. 
The gain in luminosity with this optimal choice is appreciable. 
Without optimization, i.e. working at $\sqrt{s}=500$\,GeV, 
the error would have been $1.2$\,GeV. Using the ISR corrected formulae 
one would find very similar results.
No background was assumed in the above analysis.
\end{document}